\documentclass[12pt,preprint]{aastex}
\shorttitle{Galaxy alignment and coupled Dark Energy}
\shortauthors{Baldi, Lee \& Macci\`o}
\begin{document}
\title{The Effect of Coupled Dark Energy on the Alignment between Dark Matter and 
Galaxy Distributions in Clusters}
\author{Marco Baldi\altaffilmark{1,2}, Jounghun Lee\altaffilmark{3} 
and Andrea V. Macci\`o\altaffilmark{4}}
\altaffiltext{1}{Excellence Cluster Universe, Boltzmannstr.~2, D-85748 Garching, Germany ; 
marco.baldi@universe-cluster.de}
\altaffiltext{2}{University Observatory, Ludwig-Maximillians University Munich, Scheinerstr. 1, 
D-81679 Munich, Germany}
\altaffiltext{3}{Astronomy Program, Department of Physics and Astronomy, FPRD, Seoul National 
University, Seoul 151-747, Korea ; jounghun@astro.snu.ac.kr}
\altaffiltext{4}{Max-Planck-Institut f\"ur Astronomie, K\"onigstuhl 17, 69117 Heidelberg, Germany ; 
maccio@mpia.de}

\begin{abstract}
We investigate the effects of a coupled Dark Energy (cDE) scalar field on the alignment between 
satellites  and matter distributions in galaxy clusters. Using high-resolution N-body simulations for 
$\Lambda$CDM and cDE cosmological models, we compute the probability density distribution for the 
alignment angle between the satellite galaxies and underlying matter distributions, finding a difference 
between the two scenarios. With respect to $\Lambda$CDM, in cDE cosmologies the satellite galaxies are less 
preferentially located along the major axis of the matter distribution, possibly reducing the tension with 
observational data. A physical explanation is that the coupling between dark matter and dark energy acts as
an additional tidal force on the satellite galaxies diminishing the alignments between their distribution and 
the matter one. 
Through a Wald test based on the generalized $\chi^{2}$ statistics, the null hypothesis that the two 
probability distributions come from the same parent population is rejected at the $99\%$ confidence level. 
It is concluded that the galaxy-matter alignment in clusters may provide a unique probe of dark sector interactions 
as well as the nature of dark energy. 
\end{abstract}

\section{INTRODUCTION}

The concordance cosmological model -- based on a family of non-relativistic massive particles 
possibly weakly interacting with the standard sector of particle physics (Cold Dark Matter particles, 
CDM hereafter) and on a cosmological constant $\Lambda $ -- very successfully accounts for a large wealth 
of observational data, but poses significant theoretical puzzles and requires an extreme fine-tuning of 
its basic parameters. In this context, alternative scenarios have been explored, as i.e. the possibility 
that the cosmological constant $\Lambda $ be replaced by a Dark Energy (DE) component represented by a 
classical dynamical scalar field $\phi $ evolving in a self interaction potential 
\citep[][]{Wetterich_1988,Ratra_Peebles_1988}. A further step in the exploration of alternative cosmological 
models has been recently proposed, speculating about a possible direct interaction between the DE and the 
CDM sectors of the Universe \citep[][]{Wetterich_1995,Amendola00,Amendola_2004}.
Such coupled Dark Energy models (cDE) have been widely investigated in recent years
concerning their background evolution and their effects on structure formation
\citep[e.g. by][]{Mainini_Bonometto_2006,Pettorino_Baccigalupi_2008,abdalla-etal09,Wintergerst_Pettorino_2010} by 
means of specifically designed N-body algorithms aimed at exploring the nonlinear evolution of cosmic structures 
within these alternative scenarios \citep[][]{Maccio-etal04,baldi-etal10,Baldi_2010,Li_Barrow_2010}.
These studies have led to the highlight of some distinctive features in the evolved matter density
fields of cDE models and have shown how the properties of nonlinear collapsed objects
can be significantly affected by the dark interactions in a potentially observable way 
\citep[][]{BV10,Baldi_Pettorino_2010}. Therefore, one of the most important tasks in the present investigation 
of these alternative models consists in linking the predictions obtained for the CDM density distribution to 
directly observable quantities.

\citet{lee10} very recently brought up a speculative idea, namely that the dark sector 
interaction might be probed by measuring the alignment between galaxy and matter distributions in triaxial 
clusters. In $\Lambda$CDM cosmology, the spatial correlations of the large-scale tidal fields yield the 
strong alignments of the satellite galaxies in the clusters with the cluster dark matter distributions 
\citep[e.g.,][]{altay-etal06}.  \citet{lee10} claimed that the alignment of the cluster galaxies with the underlying 
CDM distribution would be weaker in the cDE models than in $\Lambda$CDM since the spatial 
correlations of the tidal fields inside the clusters would be less strong due to the existence of a fifth force. 
If this is really the case, the alignment between galaxy and matter distributions in triaxial clusters would 
provide a unique probe of the cDE scenarios, especially because the alignment angle is insensitive to the other 
cosmological parameters. 

To back up the speculative idea of \citet{lee10},  it is essential to examine numerically whether or not the 
cDE scalar field truly makes a detectable  difference on the cluster galaxy-matter alignment.
The goal of this Paper is to use high-resolution N-body simulations to study how the alignment between 
satellites and the underlying dark matter distribution in galaxy clusters changes in cDE models as compared 
to $\Lambda$CDM \citep{wmap7}. 

\section{SIMULATIONS AND DATA ANALYSIS}

The dark interaction between the DE scalar field $\phi $ and CDM which characterizes cDE models
affects the growth of cosmic structures in a manifold way.
First of all, the background expansion history of the universe is changed with respect to the
standard $\Lambda $CDM evolution by the dynamical nature of the DE scalar field $\phi $.
Furthermore, the mass of CDM particles changes in time due to the exchange of energy
with the DE sector, and the coupling between the two fields determines a new long-range
fifth force between CDM particles and an effective ``modified inertia" in the form of
a velocity-dependent acceleration. These features can be expressed by a modified
acceleration equation for CDM particles in the Newtonian limit:
\begin{equation}
\label{acceleration_equation}
\dot{\vec{v}}_{i} = \beta (\phi )\frac{\dot{\phi }}{M_{{\rm Pl}}} \vec{v}_{i} + 
\sum_{j \ne i}\frac{G\left[ 1+2\beta ^{2}(\phi )\right] m_{j}(\phi )\vec{r}_{ij}}{|\vec{r}_{ij}|^{3}} \,,
\end{equation}
where the reduced Planck mass is defined as $M_{{\rm Pl}}\equiv 1/\sqrt{8\pi G}$, $\beta (\phi )$ is the 
coupling strength,
$\vec{v}_{i}$ is the velocity of the $i$-th particle, $\vec{r}_{ij}$ is the vector distance between the 
$i$-th and the $j$-th particles, and the sum extends to all the CDM particles in the universe. The first term 
on the right-hand side of Equation~\ref{acceleration_equation} represents the ``modified inertia", while the 
$\left[ 1+2\beta ^{2}(\phi )\right]$ factor that multiplies the gravitational constant is the ``fifth-force". 
All these effects have been studied in the literature \citep[see e.g.][]{Baldi_2010,Baldi_2010b} -- to which 
we refer the interested reader for a more detailed explanation of the cDE effects -- and need to be properly
included in N-body algorithms in order to consistently evolve the growth of density perturbations
in these scenarios. 

In the present work, we rely on the numerical implementation of cDE models in the TreePM
cosmological N-body code {\small GADGET-2} \citep[][]{gadget-2} presented in \citet{baldi-etal10}, with 
which we run a series of cosmological simulations for a fiducial $\Lambda $CDM cosmology with WMAP7 parameters
and two cDE scenarios with the same background parameters and with linear density perturbations
normalized at the present time $z=0$. Our cDE models consist of one constant coupling scenario with $\beta = 0.2$
(the EXP005 model of \citet{Baldi_Pettorino_2010}) and one variable coupling model with 
$\beta = 0.4\cdot e^{-2\phi /M_{{\rm Pl}}}$ (the EXP010e2 model of \citet{Baldi_2010}).  
Notice that the coupling definition assumed in this work is different from some of the literature and our values 
of the coupling are $\sqrt{2/3}$ smaller than the ones quoted in e.g. \citet{Amendola00}. 
For these three different cosmologies we run a CDM simulation on a cosmological box of 
$100$ comoving $h^{-1}$Mpc aside, with $512^{3}$ particles. This corresponds to a space resolution of
$\epsilon _{s} = 4\,h^{-1}$ kpc and to a mass resolution at $z=0$ of $m = 5.6 \times 10^{8}\,h^{-1}M_{\odot}$.
We then link CDM particles into groups with a friends-of-friends procedure and for each main halo we  
identify all the bound substructures by means of the {\small SUBFIND} algorithm \citep{Springel-etal01}.
Let us remind that we use the same initial random field in all our simulations. This allows us the possibility 
of a one-to-one comparison between different models.

In order to compare the results of our numerical simulations with observations we need to assign to each 
(sub)halo a stellar mass. For this purpose we use the recently derived stellar-to-halo mass relationship 
presented in \citet{Moster-etal10}. In this paper the authors give a series of simple fitting formulae to 
derive the stellar mass of the galaxy hosted by a given dark matter halo as a function 
of the virial mass (for central galaxies) or as a function of the maximum mass over the halo's history 
(for satellite galaxies). For all our substructures we therefore build a full merger tree and we assign to 
each subhalo a stellar mass based on the latter criterion by using the formulae of \citet{Moster-etal10}.

\begin{figure}
\includegraphics[width=1.0\columnwidth]{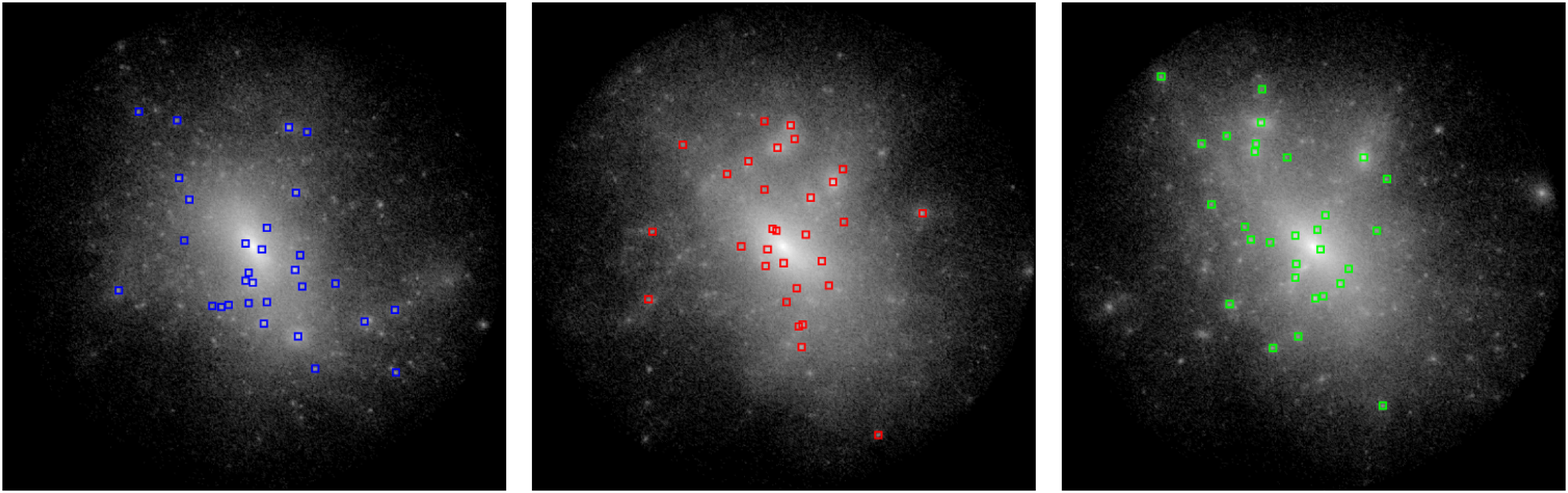}
\caption{CDM projected density (grey scale) and location of the 30 most luminous substructures (colored squares) within $R_{200}$ for the most massive cluster formed in our simulated $\Lambda $CDM (left), EXP005 (middle) and EXP010e2 (right) cosmologies, respectively.}
\label{fig:halos}
\end{figure}

However, these formulae have been derived by comparing the observed galaxy luminosity function with the halo 
mass function in a $\Lambda$CDM universe. This implies that we cannot directly apply them in the case of 
cDE models. In order to overcome this issue, we first find a function $f(M,z)$ able to map  the cDE 
mass function onto the $\Lambda$CDM one and then we apply the \citet{Moster-etal10} fitting equations. 
The CDM density and the location of the 30 most luminous substructures  inside two virial radii for the same 
massive cluster within the three different cosmologies under investigation are plotted in the three panels of 
Figure~\ref{fig:halos}, where a stronger tendency of alignment between the CDM distribution and the selected 
substructures in the $\Lambda $CDM model as compared to cDE scenarios is already visible. 

\section{ALIGNMENT BETWEEN GALAXY AND MATTER DISTRIBUTIONS IN CLUSTERS}

From each cluster catalog obtained from the simulations for the $\Lambda$CDM and cDE cosmologies, we select 
those cluster halos which have five or more galaxies within some cut-off radius $r_{c} (\le R_{vir})$ from the 
halo center. Then, we measure the alignment angles between the galaxy and dark matter distributions of each 
selected cluster halo through the following procedure. First, the inertia tensor of the galaxy distribution of 
each selected cluster halo, $(I^{G}_{ij})$, is evaluated as
\begin{equation}
\label{eqn:inertia_g}
I^{G}_{ij}=\sum_{\beta}m_{*,\beta}x_{\beta,i}x_{\beta,j},
\end{equation}
where $m_{*,\beta}$ is the stellar mass of the $\beta$-th galaxy and $(x_{\beta, i})$ is the position vector of 
the $\beta$-th satellite galaxy relative to the center of the stellar mass of the galaxy distribution. Weighting the 
satellites by stellar mass is important for a comparison with observations since in practice the shapes of galaxy 
distributions are often obtained from the luminosity-weighted satellites. Finding the unit eigenvector of $(I^{G}_{ij})$ 
corresponding to the largest eigenvalue, we determine the major axis, ${\bf e}_{G}$, of the galaxy distribution of each 
selected cluster halo.

\begin{figure}
\includegraphics[width=\columnwidth]{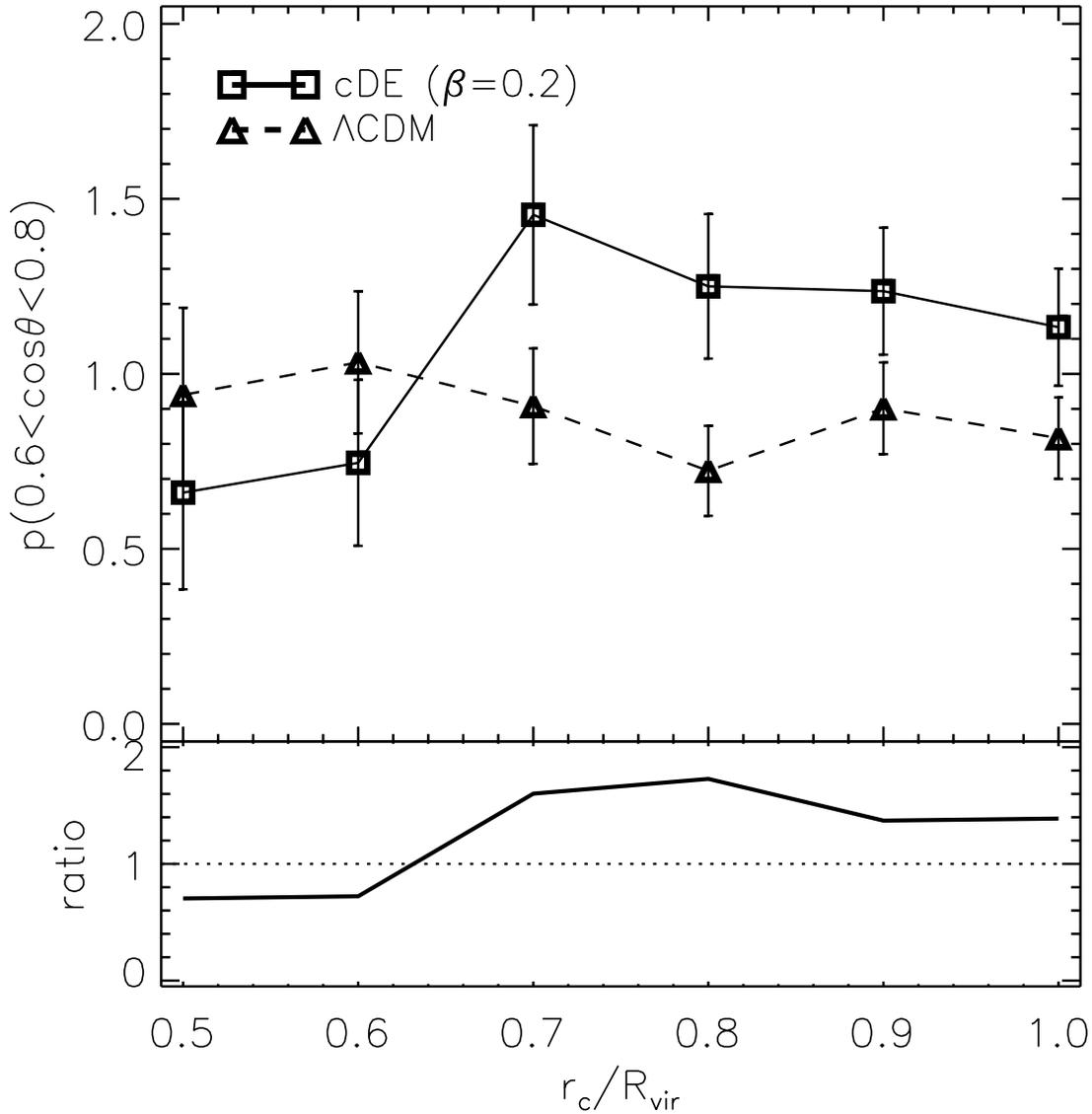}
\caption{Probability  density that the cosines of the angles between the major axes of dark matter and 
galaxy distributions of the simulated clusters are in the range of $(0.6, 0.8)$ with the bootstrap errors as 
a function of the cut-off radius of the satellite galaxies $r_{c}$ for two cosmological models: cDE (solid line+
square dots) and $\Lambda$CDM (dashed line+triangle dots). in the top panel 
The ratios of the square points to the triangle points at each bin are plotted as solid line in the bottom panel.}
\label{fig:ar_align}
\end{figure}

Similarly, we evaluate the inertia tensor, $(I^{M}_{ij})$, of the CDM distribution of each selected cluster halo as
\begin{equation}
\label{eqn:inertia_dm}
I_{ij}^{M}=\sum_{\alpha}x_{\alpha,i}x_{\alpha,j},
\end{equation}
where $(x_{\alpha,i})$ is the position vector of the $\alpha$-th dark matter particle relative to the 
halo center.  The summation  in Equation (\ref{eqn:inertia_dm}) goes over all dark matter particles 
located within the halo virial radius $R_{vir}$. Finding the unit eigenvector of $(I^{M}_{ij})$ corresponding 
to the largest eigenvalue, we determine the major axis, ${\bf e}_{M}$, of the CDM distribution of each 
selected cluster halo. The alignment angle, $\theta$, between dark matter and galaxy distributions of each 
selected halo is now calculated as $\cos\theta=\vert{\bf e}_{M}\cdot{\bf e}_{G}\vert$ 
where $\cos\theta$ is in the range of $[0,1]$. Binning the values of $\cos\theta$ and counting the number of 
the halos belonging to each $\cos\theta$-bin, we derive the probability density distribution, $p(\cos\theta)$, 
for the $\Lambda$CDM and cDE cosmologies, separately.  When the cut-off radius $r_{c}$ for 
the satellite galaxies is set at $0.8R_{vir}$ (see Figure \ref{fig:ar_align}), the total number of the selected cluster 
halos which have five or more galaxies within $r_{c}$ is found to be $186$ and $116$ for the $\Lambda$CDM and 
cDE cosmologies, respectively.

\begin{figure}
\includegraphics[width=\columnwidth]{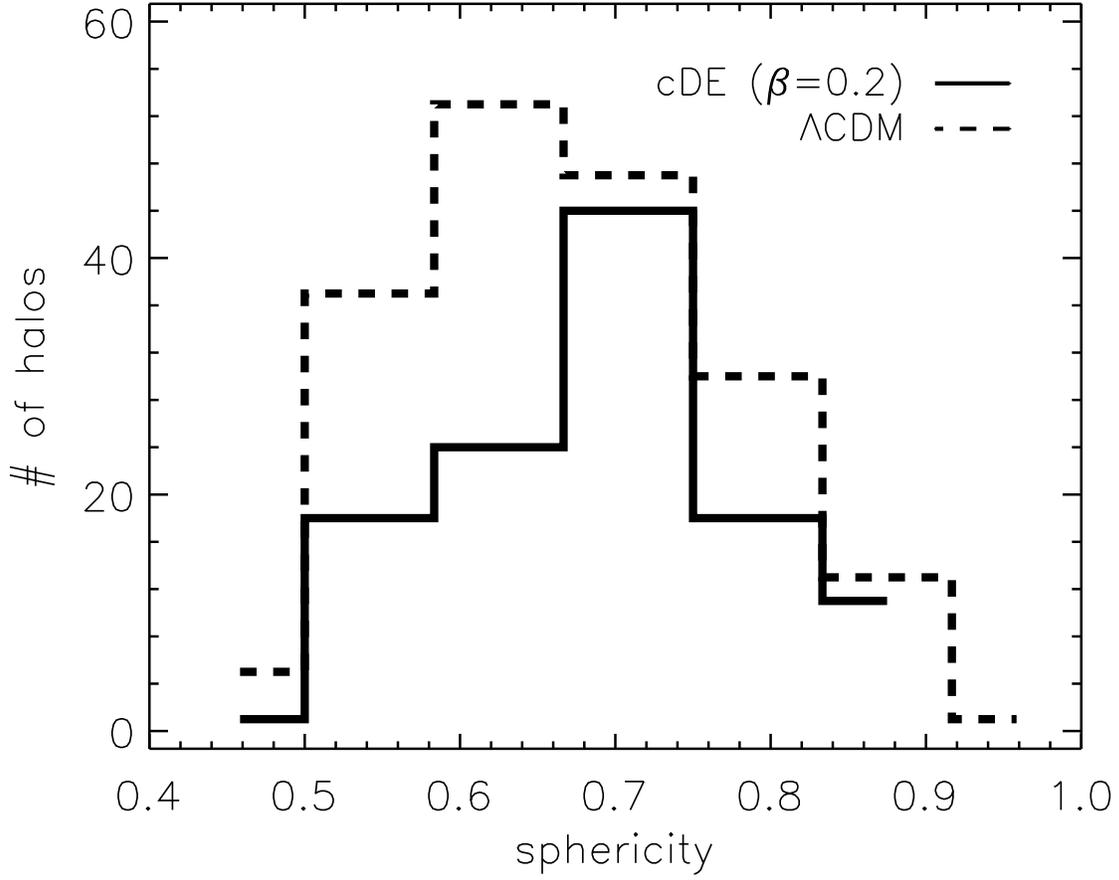}
\caption{Number distributions of the selected halos as a function of the sphericity of their dark matter distribution 
for two different cosmologies:cDE (solid histogram) and $\Lambda$CDM (dashed histogram).}
\label{fig:shape}
\end{figure}

Before making a comparison between the probability distributions of $\cos\theta$ from the two cosmologies, 
it is worth examining how robust our measurement of $\cos\theta$ is.  If the dark matter distribution of a given 
halo is spherically symmetric, then the determination of the major principal axis of its inertia tensor would be 
quite ambiguous. Therefore, the robustness of the measurement of $\cos\theta$ of a given halo critically depends 
on how spherical its dark matter distribution is. The sphericity of a given halo is conventionally defined as 
$S\equiv \sqrt{u_{3}/u_{1}}$, where $u_{1}$ and $u_{3}$ are the largest and the smallest eigenvalues of 
the inertia tensor of the dark matter distribution, respectively.  Using this definition, we measure the sphericity of each 
selected halo for both cosmologies. Figure \ref{fig:shape} plots the number distribution of the selected halos 
versus their sphericity for cDE and $\Lambda$CDM cosmologies as solid and dashed histogram, respectively. 
As it can be seen, for the case of cDE there is no halo whose sphericity is larger than $0.9$, 
while for  $\Lambda$CDM there is only one halo whose sphericity exceeds 
$0.9$. Thus, our measurement of $\cos\theta$ is robust, free from the ambiguity caused by the spherically 
symmetric dark matter distribution. 

\begin{figure}
\includegraphics[width=\columnwidth]{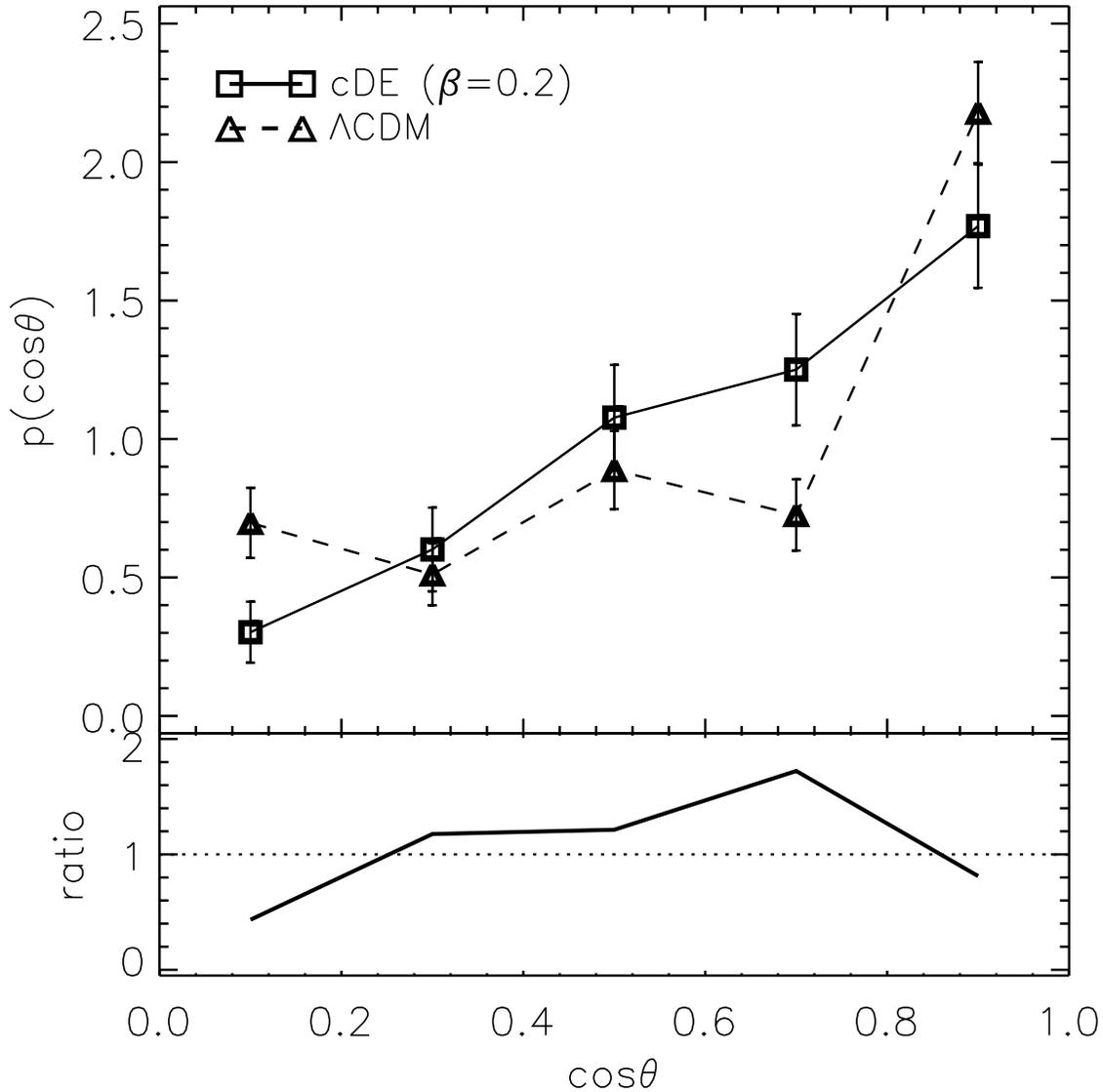}
\caption{Probability density distribution of the cosines of the angles between the major axes 
of dark matter and galaxy distributions of the simulated clusters for two different cosmologies:
cDE (solid line+square dots) and $\Lambda$CDM (dashed line+triangle dots) in the top panel. 
The errors represent the $1\sigma$ scatter among the $1000$ bootstrap resamples.
The ratios of the square points to the triangle points are plotted as solid line in the bottom panel.}
\label{fig:3dalign}
\end{figure}

Given that our samples of the halos for both cosmologies are not large enough to be fully representative of the 
true parent populations, we perform the bootstrap error analysis to estimate the uncertainties in the measurement 
of $p(\cos\theta)$ \citep{WJ03}.  We first construct $1000$ bootstrap resamples and measure 
$p(\cos\theta)$ from each resample. The bootstrap error, $\sigma_{b}$, at each bin is then calculated as 
one standard deviation among the $1000$ remeasurements,  
$\sigma_{b}\equiv\langle[p(\cos\theta_{i})-\langle p(\cos\theta_{i})\rangle]^{2}\rangle^{1/2}$, where 
the ensemble average is taken over the $1000$ resamples and $\langle p(\cos\theta_{i})\rangle$ represents 
the bootstrap mean value at the $i$-th bin.  The top panel of Figure \ref{fig:3dalign} plots the resulting 
$p(\cos\theta)$ with the bootstrap errors for cDE and $\Lambda$CDM cosmologies as square and 
triangle dots, respectively. For this plot, the cut-off radius $r_{c}$ for the satellite galaxies is set at $0.8R_{vir}$.
 
As shown in Figure \ref{fig:3dalign}, the probability density distribution increases with $\cos\theta$ in both 
cosmologies, which indicates that the major axes of the galaxy and dark matter distributions in clusters tend to be 
strongly aligned. However,  we note a difference between the two cosmologies: 
$p(\cos\theta)$ increases less rapidly in the cDE model. At the fourth $\cos\theta$-bin 
(corresponding to $0.6\le\cos\theta\le 0.8$), the difference between the two models exceeds $3\sigma_{b}$.
The bottom panel of Figure \ref{fig:3dalign} plots the ratio between the two probability density distributions: 
$p_{cDE}(\cos\theta)/p_{\Lambda CDM}(\cos\theta)$, which demonstrates clearly that the ratio deviates 
from unity and the degree of the deviation reaches the maximum at the fourth bin. 

Now that we find a difference between the two cosmologies in the tendency of the cluster galaxy-matter 
alignment, we would like to investigate how the result depends on the cut-off radius $r_{c}$. 
In fact, it is naturally expected that as $r_{c}$ decreases the alignment tendency would decrease. 
The major axes of the dark matter distribution are determined at $R_{vir}$ while the major axes of the galaxy 
distribution are measured at $r_{c}$.  Therefore, the difference $R_{vir}-r_{c}$ represents the distance 
by which the inner and outer tidal fields are separated. As $r_{c}$ decreases, the spatial correlations 
between the two tidal fields would decrease, which would be manifested by the decrease of the strength of 
the alignment between the major axes of the galaxy and matter distributions in clusters.

Repeating the whole process described above, we calculate the probability density at the fourth bin, 
$p(0.6<\cos\theta< 0.8)$, for five different choices for $r_{c}$. The top panel of Figure \ref{fig:ar_align} plots 
the result for cDE and $\Lambda$CDM cosmologies as square and triangle dots, respectively. 
The bottom panel of Figure \ref{fig:ar_align} plots the ratio of the square points (cDE) to the triangle points 
($\Lambda$CDM). Figure~\ref{fig:ar_align} reveals  that in the cDE scenario the probability density, 
$p(0.6<\cos\theta<0.8)$, decreases more rapidly as $r_{c}$ decreases. This result may be interpreted as the 
effect of coupled dark energy on the spatial correlations of the tidal fields inside the cluster halos. 
Due to the coupling between dark matter and dark energy, there is an additional tidal force exerted on the 
galaxies inside the clusters, which plays a role of reducing further the correlations of galaxy distributions 
with matter distributions. Furthermore, the velocity-dependent acceleration arising as a consequence of 
the ``modified inertia" described by the first term on the right-hand side of Equation~(\ref{acceleration_equation})
also contributes to an effective modification of the internal tidal field of the cluster 
experienced by the substructures since it determines an effective extra force along the direction of motion 
of the satellite halos.

\begin{figure}
\includegraphics[width=\columnwidth]{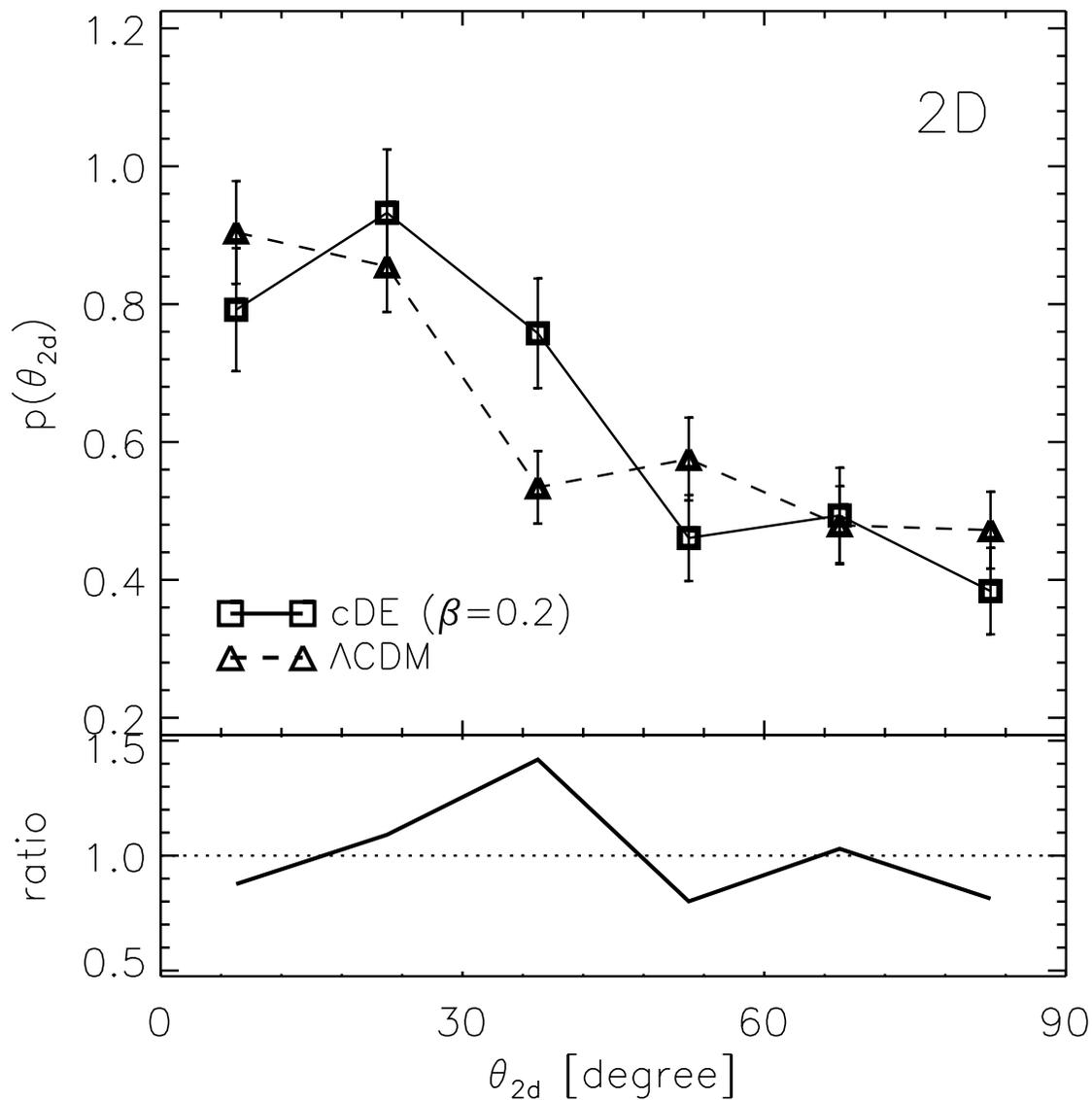}
\caption{Probability density distribution of the angles between the two dimensional projected major axes 
of dark matter and galaxy distributions with the bootstrap errors for two different cosmologies:
cDE (solid line+square dots) and $\Lambda$CDM (dashed line+triangle dots). The ratios  
of the square points to the triangle points are plotted as solid line in the bottom panel.}
\label{fig:2dalign}
\end{figure}

To make a direct comparison with observations, it may be more useful to find the probability density 
distribution of the two dimensional projected alignment angles, $\theta_{\rm 2d}$. For each cosmological 
model, we project ${\bf e}_{G}$ and ${\bf e}_{M}$ of each  selected halo cluster onto the xy, yz and zx planes 
and calculate the two dimensional alignment angle $\theta_{\rm 2d}$ between the projected major axes of 
dark matter and galaxy distributions for all three projection cases and determine the probability density 
distribution, $p(\theta_{2d})$, by taking the average over the three cases. To estimate the uncertainties 
in the measurement of $p(\theta_{2d})$, we also perform the boostrap error analysis. 
The top panel of Figure \ref{fig:2dalign} plots the $p(\theta_{2d})$ with the bootstrap errors for cDE and 
$\Lambda$CDM cosmologies as square  and triangle dots, respectively. As one can see, the two dimensional 
projection does not dilute the difference in the alignment tendency between the two models. The probability density 
distribution, $p(\theta_{\rm 2d})$, decreases less rapidly as $\theta_{\rm 2d}$ increases in the cDE model 
than in the $\Lambda$CDM model. Note that for a three dimensional distribution, it is $p(\cos\theta)$ that is expected to be uniformly distributed in case of no alignment while for the two dimensional case, 
it is $p(\theta_{\rm 2d})$.

To quantify the statistical significance of the difference in the galaxy-matter alignments between 
$\Lambda$CDM and cDE cosmologies, we perform a Wald test for the ratio, 
$\xi\equiv p_{cDE}(\cos\theta)/p_{cDE \Lambda CDM}(\cos\theta)$. 
In the null hypothesis $H_{N}$ that the two probability density distributions shown in Figures   
\ref{fig:3dalign} and \ref{fig:2dalign} are in fact from the same parent population,  the expectation value of 
$\xi$ is unity at all bins. Taking into account the possible cross-correlations of $p(\cos\theta)$ between the 
different bins, we calculate the generalized $\chi^{2}$ defined in terms of the full covariance matrix as
\begin{equation}
\label{eqn:chi2}
\chi^{2}\equiv (\xi_{i} - 1)C^{-1}_{ij}(\xi_{j} - 1).
\end{equation}
Here $C^{-1}_{ij}$ is the inverse of the covariance matrix, 
$C_{ij}\equiv \langle(\xi_{i}-\bar{\xi}_{i})(\xi_{j}-\bar{\xi}_{j})\rangle$, where the ensemble average is 
taken over the $1000$ bootstrap resamples and $\bar{\xi}_{i}$ denotes the mean value of $\xi$ averaged 
over the bootstrap resamples at the $i$-th bin. In Equation (\ref{eqn:chi2}), $(\xi_{i}-1)$ expresses how 
much the numerical result of $\xi$ at the $i$-th bin deviates from the expectation value of unity. 
We find $\chi^{2}=15.08$ and $14.97$ for the three and two dimensional case, respectively. 
Since the number of degrees of freedom for $\chi^{2}$ is $5$, its value corresponds to the null hypothesis being 
rejected at the $98.9\%$ confidence level for both cases.

\section{DISCUSSION AND CONCLUSIONS}

Using data from high-resolution N-body simulations, we have shown that the $\Lambda$CDM and cDE
cosmologies yield different strengths of the alignment between satellite galaxy and matter distributions in 
triaxial clusters. As speculated by \citet{lee10}, it is found that the alignment is less strong in cDE 
cosmologies than in the standard $\Lambda$CDM scenario.  The null hypothesis that the two cosmologies give 
the same alignment tendency is rejected at the $98.9\%$ confidence level through a Wald test based on the 
generalized $\chi^{2}$ statistics. 

The differences in the satellite distributions between $\Lambda$CDM and cDE models may be understood as 
follows.  There are two possible mechanisms for explaining the alignments between the dark matter and galaxy 
distributions \citep{altay-etal06}.  The first one is the spatial correlations between the large scale external tidal 
fields and the internal tidal fields due to the cluster potentials which affect he non-spherical shapes of dark 
matter components of galaxy clusters and the anisotropic distributions of their satellite galaxies, respectively. 
The second one is the filamentary merging/accretion of matter and galaxies along the large-scale filaments, 
which also induce the alignments between the dark matter and galaxy distributions in clusters. 

In the $\Lambda$CDM cosmology, the tidal fields evolve nonlinearly only via gravity.  
Whereas in the cDE cosmology the coupling between dark energy and dark matter generate additional 
(more isotropic) tidal effects in the clusters, which weaken the spatial correlations between the external and 
internal tidal fields and also redistribute previously accreted satellites that were initially in a flattened distribution. 
Furthermore, the ``modified inertia" characterizing cDE models would contribute to an effective enhancement 
of the internal cluster tidal field. 

Recently \citet{oguri-etal10} compared the observed distribution of satellite galaxies in clusters with the
shape of the cluster dark matter distribution determined from weak lensing, finding a very weak 
alignment \citep{lee10}. Due to the small sample (composed of only $13$ clusters), it is difficult to say at 
the moment whether or not this observational signal supports the cDE scenario. In any case, a crucial 
implication of our results is that the alignment between galaxy and matter distributions in clusters is in 
principle a unique probe of dark sector interactions.  

\acknowledgments

We thank an anonymous referee for helpful suggestions.
M.B. is supported by the DFG Cluster of Excellence ``Origin and Structure of the Universe''.
J.L. acknowledges the support from the National Research Foundation of Korea (NRF) grant funded by 
the Korea government (MEST, No.2010-0007819) and partial support provided by the National Research Foundation 
of Korea to the Center for Galaxy Evolution Research.
All the numerical simulations have been performed on the VIP cluster at the RZG computing center in Garching.

\clearpage
\end{document}